\documentclass[12pt]{iopart}
\usepackage{iopams}
\usepackage{citesort}
\usepackage[dvips]{graphicx}
\usepackage{xcolor}
\usepackage{mathrsfs}
\usepackage{subeqn}

\newcommand{\eqind}{\,{\buildrel d \over =}\,}

\begin{document}
\title[]{Accelerating and Retarding Anomalous Diffusion}

\author{Chai Hok Eab}
\address{%
Department of Chemistry
Faculty of Science, Chulalongkorn University
Bangkok 10330, Thailand%
}
\ead{Chaihok.E@chula.ac.th}

\author{S.C. Lim}
\address{%
Faculty of Engineering, Multimedia University
63100 Cyberjaya, Selangor Darul Ehsan, Malaysia%
}
\ead{sclim47@gmail.com}

\begin{abstract}
In this paper Gaussian models of retarded and accelerated anomalous diffusion are considered. 
Stochastic differential equations of fractional order driven by single or multiple fractional Gaussian noise terms are introduced to describe retarding and accelerating subdiffusion and superdiffusion. 
Short and long time asymptotic limits of the mean squared displacement of the stochastic processes associated with the solutions of these equations are studied.
Specific cases of these equations are shown to provide possible descriptions of retarding or accelerating anomalous diffusion.
\end{abstract}


\maketitle

\section{Introduction}
\label{sec:ADintro}
Anomalous diffusion occurs in many physical, chemical and biological systems
\cite{MetzlerKlafter00,KlagesRadonsSokolov08,KlafterLimMetzler11}. 
In normal diffusion the mean squared displacement of diffusion particles varies linearly with time $\bigl<\Delta{x^2(t)}\bigr> \sim t$.
In some complex disordered media the diffusion becomes anomalous with $\bigl<\Delta{x^2(t)}\bigr> \sim t^\alpha$, where the scaling exponent $\alpha \neq 1$ characterizes the anomalous diffusion.
For $\alpha < 1$ the process is known as subdiffusion, and when $\alpha > 1$ it is called superdiffusion.
Differential equations of fractional order are well-suited for describing fractal phenomena such as anomalous diffusion in complex disordered media.
The constant memory and self-similar character of these phenomena can be taken into account by using the kinetic equations with fixed fractional order.

There exist in certain locally heterogeneous media where diffusing processes do not satisfy the constant power-law type scaling behavior like anomalous diffusion. 
Such processes include the retarding and accelerating anomalous diffusion. Retardation of diffusion occurs in single-file diffusion where particles are constrained to move in single-file due to confined one-dimensional geometry, such as diffusion in zeolites \cite{KargerRuthven92}, 
or in the anomalous diffusion that occurs on the biological cell membrane 
\cite{Saxton96,WeissElsnerKartbergNilsson04}. 
Possible causes for the presence of anomalous subdiffusion in biological systems include the presence of immobilized obstacles which hamper molecular motion by an excluded volume interaction, and the cytoplasmic crowding in living cells.
On the other hand, membrane-bound proteins exhibit transition from subdiffusion at short time to normal or superdiffusion at long times 
\cite{KhanReynoldsMorrisonCherry05}; 
the diffusion of telomeres in the nucleus of mammalian cells shows accelerating subdiffusion 
\cite{Bronstein09}. 
Other examples are retarded and enhanced dopant diffusion in semiconductors 
\cite{BergholzHutchisonPirouz85,Agarwal99,TanGosele05}, 
the accelerating superdiffusion of energetic charged particles across magnetic field in astrophysical plasma physics 
\cite{KirkDuffyGallant96,PerriZimbardo07}, 
and accelerated diffusion in Josephson junction 
\cite{GeiselNierwetbergZacheri85}. 
Such processes have memory and fractal dimension that may vary with position, temperature, density, 
or internal parameters such as elasticity and viscosity.

Anomalous diffusion having scaling exponent which varies with position and time was studied by Glimm and co-workers in the early 1990s 
\cite{Furtado92,Glimm93} 
in the multifractal modeling of heterogeneous geological systems.
More systematic studies of transport phenomena with variable scaling exponents were carried out in the late 1990s and early 2000s. 
Stochastic processes with variable fractional order such as multifractional Brownian motion was introduced to model phenomena with variable memory or variable fractal dimension 
\cite{PeiltierLevyVehel96,BenassiJaffardRoux97,Lim01}. 
In order to describe systems with variable scaling exponents one may have to consider fractional differential equations of variable order. 
However, such variable order equations are in general mathematically intractable and can not be solved without numerical approximations 
\cite{RossSamko93,Samko95,LorenzoHartley02,KobelevKobelevKlimontovich03,ChechkinGorenfloSokolov05,SunChenChen09}.

There exists certain class of diffusion processes with non-unique scaling exponent that can be described by fractional differential equations of distributed order.
The notion of distributed order differential operators was first introduced by Caputo in 1969 
\cite{Caputo67}. 
A distributed-order fractional diffusion equation has its fractional order derivatives integrated over the order of differentiation within a given range. 
Applications of distributed fractional order equations to fractional diffusion and fractional relaxation have been carried out by various authors 
\cite{%
ChechkinGorenfloSokolov05,SunChenChen09,Caputo67,Caputo01,%
ChechkinGorenfloSokolov02,ChechkinGorenfloSokolovGonchar02,%
MeerschaertScheffler06,MainardiMuraPagniniGorenflo08,EabLim11%
}.

This paper considers Gaussian models of retarded and accelerated anomalous diffusion. 
We introduce a class of multi-term fractional Langevin-type equations driven by single or multiple fractional Gaussian noise terms. 
These equations can be regarded as special cases of fractional Langevin equations of distributed order, and they can be used to describe retarded and accelerated anomalous diffusion. 
Detailed study of the short and long time asymptotic properties of the solutions to these equations are carried out.

\section{Multi-fractional stochastic differential equations}
\label{sec:multifrac}
In this section we introduce a class of multi-fractional Langevin-like equations of the following form:
\begin{equation}
  \sum_{i=1}^m a_iD^{\alpha_i}x(t)  =  \sum_{j=1}^n c_j\xi_{\gamma_j}, \qquad
                               t \in \mathbb{R},  0 < \alpha_i \leq 2
\label{eq:sec:multifrac_0010}
\end{equation}
where $c_j > 0$, $D_t^{\alpha_i}$ is the Riemann-Liouville or Caputo fractional derivative 
\cite{SamkoKilbasMaritchev93,Podlubny99,WestBolognaGrigolini03,%
KilbasSrivastavaTrujillo06,Mainardi10}
which is defined for $m-1 \leq \alpha \leq m$ as
\begin{equation}
\fl\qquad
  {D_t^\alpha} f(t) = \left\{
    \begin{array}{ll}
      \displaystyle
      \frac{1}{\Gamma(m-\alpha)} \frac{\rmd^m}{\rmd t^m}\int_0^t \frac{f(u)du}{(t-u)^{\alpha-m+1}}, 
          & \textrm{Riemann-Liouville} \\
      \displaystyle
      \frac{1}{\Gamma(m-\alpha)} \int_0^t (t-u)^{m-\alpha-1}\frac{\rmd^m}{\rmd t^m}f(u)du, 
          & \textrm{Caputo}
    \end{array}
    \right. .
\label{eq:sec:multifrac_0020}
\end{equation}
The Gaussian noise $\xi_{\gamma_j}(t)$ is defined by
\begin{equation}
  \Bigl<\xi_{\gamma_j}(t)\Bigr> = 0 ,
\label{eq:sec:multifrac_0030}
\end{equation}
and
\begin{equation}
  \Bigl<\xi_{\gamma_i}(t)\xi_{\gamma_j}(s)\Bigr> = \delta_{ij}d_j|t - s|^{-\gamma_j}, 
  \qquad 0 < \gamma_i, \gamma_j < 2 ,
\label{eq:sec:multifrac_0040}
\end{equation}
with
\begin{equation}
  d_j = \frac{1}{2\sin(\pi\gamma_j/2)\Gamma(1-\gamma_j)}.
\label{eq:sec:multifrac_0050}
\end{equation}
If we let $\gamma_j = 2 -2H_j$, where $0 < H_j < 1$ is the Hurst index associated with fractional Brownian motion, 
and $d_j = (2\sin(\pi H_j)\Gamma(2H_j -1))^{-1}$.  
$\xi_{\gamma_j}(t)$ can then be regarded as the derivative (in the sense of generalized functions) of fractional Brownian motion indexed by $H_j$.
Note that the covariance of fractional Gaussian noise has the same algebraic sign as $(2H_j -1)$.
For $1/2 \leq H_j < 1$, the process exhibits long-range dependence with persistent positive covariance.
On the other hand, when $0<H_j < 1/2$, $d_j$ is negative and the process has anti-persistent correlation structure.
For $H_j = 1/2$, or $\gamma_j =1$, it corresponds to white noise. 
Here we remark that $\lim_{H_j \to 1/2} d_j|t|^{2H_j-2} = \delta(t)$ in the sense of generalized functions 
\cite{GelfandI,GelfandIV}.
We thus see that there is no need to include in the covariance of fractional Gaussian noise $\xi(t)$ an extra term to cater for the white noise when $H = 1/2$ as given by 
\cite{Qian03} 
\begin{equation}\fl\qquad
  \Bigl<`\xi_{H_j}(t)\xi_{H_j}(s)\Bigr> = 4d_jH_j(2H_j -1 )|t-s|^{2H_j-2} + 4d_jH_j|t-s|^{2H_j-1}\delta(t-s).
\label{eq:sec:multifrac_0060}
\end{equation}

Here we would like to briefly discuss the covariance of fractional Gaussian noise 
\cite{Qian03,BartonPoor88,Zinde-WalshPhillips03} 
in terms of generalized functions.
Just like for a proper description of fractional Gaussian noise, $\xi(t)$ should not be defined pointwise for each $t$.
Instead it needs to be considered the process as a linear functional in some test function space such as Schwarz space 
$\mathscr{S}(\mathbb{R})$
of real-valued infinitely differentiable functions which decrease rapidly 
\cite{GelfandI}. 
The generalized process $\xi(t)$, $f\in \mathscr{S}(\mathbb{R})$, is a linear functional
\begin{equation}
  \xi(f)  = \int_{-\infty}^\infty \xi(t)f(t)dt.
\label{eq:sec:multifrac_0070}
\end{equation}
$\xi(f)$ is a generalized stationary Gaussian process with covariance given by the bilinear functional
\begin{eqnarray}
  C(f,g) & = \Bigl<\xi(f)\xi(g)\Bigr> 
           = \int_{-\infty}^\infty\int_{-\infty}^\infty f(t)g(s)c(t-s)dtds \nonumber \\
         & =  \int_{-\infty}^\infty f(t)dt\int_{-\infty}^\infty g(s)c(|t-s|)ds \nonumber \\
         & = \int_{-\infty}^\infty f(t)dt\int_0^\infty c(s)\Bigl[g(t+s) + g(t-s)\Bigr]ds ,
\label{eq:sec:multifrac_0080}
\end{eqnarray}
where $c(t-s)$ is a generalized function or distribution.
For example, for $H =1/2$, $\xi$ is white noise with $c(t-s) = \delta(t-s)$.
$H \neq 1/2$ corresponds to fractional Gaussian noise, with $c(t)$, $t > 0$ given by
\begin{equation}\fl \qquad
  c(t) = \frac{t^{2H-2}}{\sin(\pi H)\Gamma(2H-1)} =
  \left\{
    \begin{array}{ll} \displaystyle
      \frac{1}{\sin(\pi H)} I^{2H-1} \delta(t), & 1/2 < H < 1 \\
      \displaystyle
      \frac{1}{\sin(\pi H)} D^{1-2H} \delta(t), & 0 < H < 1/2
    \end{array}
  \right.,
\label{eq:sec:multifrac_0090}
\end{equation}
When $1/2 < H < 1$ the covariance kernel in (\ref{eq:sec:multifrac_0080}) 
can be regarded as the fractional integral of delta function,
$I^{2H-1} \delta(t)$ up to a multiplicative constant $(2\sin(\pi H)^{-1}$.
In this case, $c(t-s)$ is locally integrable. 
It becomes fractional derivative of delta function for $(2\sin(\pi H)^{-1} D^{1-2H} \delta(t)$ for $0 < H < 1/2$
since $I^\alpha f = D^{-\alpha} f$
\cite{Zinde-WalshPhillips03}. 

Before we proceed further, a brief comment on the stochastic differential equation driven by fractional Gaussian noise will be given. 
Recall that stochastic calculus of Ito can not be used to define the integrals with respect to a stochastic process which is not a semimartingale. 
Fractional Brownian motion is not a semimartingale when the Hurst index of the process $H \neq 1/2$, 
or when it is not a Brownian motion. Due to its widespread application, the question on how to obtain a well-defined stochastic integral with respect to fractional Brownian motion has become a long standing problem which has attracted considerable attention 
\cite{Mushura08,BiaginiHuOksendal08}. 
Several methods which include Sokorohod-Stratonovich stochastic integrals, Malliavin calculus, and pathwise stochastic calculus have been proposed to overcome this difficulty 
(see references \cite{BiaginiHuOksendal08,AzmoodehTikanmakiValkeila10} 
for details).
However, for application purposes, theory based on abstract integrals may encounter difficulty in physical interpretations.
As we shall restrict our discussion related to applications fractional Gaussian noise involving only persistent case
$1/2 < H_j < 1$ (or $0 < \gamma_j < 1$),
the integrals with respect to fractional Brownian motion can thus be treated as the pathwise Riemann-Stieltjes integrals 
(see for example \cite{AzmoodehTikanmakiValkeila10} 
and references given there).
This allows one to handle such integrals in a similar way as ordinary integrals.

Here we remark that the multi-term fractional order Langevin-like equation 
(\ref{eq:sec:multifrac_0010}) 
can also be regarded as a special class of the following distributed-order fractional time stochastic equation:
\begin{equation}
   D_\varphi x(t) = \xi_\psi(t), \qquad t \geq 0.
\label{eq:sec:multifrac_0100}
\end{equation}
with the distributed fractional derivative
\begin{equation}
  D_\varphi x(t) = \int_0^2 \varphi(\alpha) D_t^\alpha x(t) d\alpha ,
\label{eq:sec:multifrac_0110}
\end{equation}
where $D_t^\alpha$ is the fractional derivative as defined by (\ref{eq:sec:multifrac_0020})
, the weight function $\varphi(\alpha)$, $0 \leq \alpha \leq 2$,
which satisfies $\varphi(\alpha) \geq 0$ is given by $\varphi(\alpha) = \sum_{i=1}^m a_i\delta(\alpha -\alpha_i)$.
Note that in general the weight function is a positive generalized function.
The Gaussian noise $\xi_\psi(t)$ is the distributed-order fractional Gaussian noise defined by
\begin{equation}
  \xi_\psi (t) = \int_0^2 \xi_\gamma(t)\psi(\gamma) d\gamma,
\label{eq:sec:multifrac_0120}
\end{equation}
with the weight function $\psi(\gamma)$, $ 0 \leq \gamma \leq 2$, 
which satisfies $\psi(\gamma) \geq 0$ and is given by $\psi(\gamma) = \sum_{j=1}^n c_j \delta(\gamma -\gamma_j)$.
Here we remark that in most of the examples considered subsequently, we shall restrict $0 \leq \gamma \leq 1$.

Solution of (\ref{eq:sec:multifrac_0010}) 
can be solved formally by using Laplace transform method which gives
\begin{equation}
  A(s)\tilde{x}(t) - B(s) = \tilde{\xi}(s),
\label{eq:sec:multifrac_0130}
\end{equation}
where $\tilde{x}(s)$ is the Laplace transform of $x(t)$ and
\begin{equation}
  A(s) = \sum_{i=1}^m a_i s^{\alpha_i},
\label{eq:sec:multifrac_0140}
\end{equation}
and
\begin{equation}\fl \qquad
  B(s) = \left\{\!
  \begin{array}{ll} 
    \displaystyle
    \sum_{i=1}^m a_i \sum_{k_i=0}^{\lceil{\alpha_i}\rceil} s^k\Bigl[D_{RL}^{\alpha_i-k_i-1}x(t)\Bigr]_{t=0} ,   & \textrm{(Riemann-Liouville)} \\
    \displaystyle
    \sum_{i=1}^m a_i \sum_{k_i=0}^{\lceil{\alpha_i}\rceil} s^{\alpha_i-k_i -1} x^{(k_i)}(0) ,                  &  \textrm{(Caputo)}
  \end{array}
  \right. .
\label{eq:sec:multifrac_0150}
\end{equation}
Here $\lceil{\alpha_i} \rceil$ denotes the largest integer smaller or equal to $\alpha_i$.
For Riemann-Liouville case, $\Bigl[D_{RL}^{\alpha_i-k_i-1}x(t)\Bigr]_{t=0}$ is the Riemann-Liouville derivative of order $\alpha_i-k_i-1$ evaluate at $t=0$.
For the Caputo case, $x^{(k_i)}(t)$ denotes $k_i$th derivative of $x(t)$.
For simplicity we assume the initial conditions $x^{(k_i)}(0) = 0$ for all $i = 1, \cdots, m$,
and $\Bigl[D_{RL}^{\alpha_i-k_i-1}x(t)\Bigr]_{t=0} = 0$,
such that $B(s) = 0$ for both these cases.
The Laplace transform of the Green function is then given by $\widetilde{G}(s) = 1/A(s)$.
Therefore, 
\begin{equation}
  \tilde{x}(s) = \widetilde{G}(s)\tilde{\xi}(s) = \frac{\tilde{\xi}(s)}{A(s)}.
\label{eq:sec:multifrac_0160}
\end{equation}
The solution is then given by the inverse Laplace transform:
\begin{equation}
  x(t) = \int_0^t G(t-u)\xi(u)du.
\label{eq:sec:multifrac_0170}
\end{equation}
The covariance and variance of the process are given respectively by
\begin{equation}
  K(s,t) = \int_0^tdu\int_0^sdv G(t-u)C(u-v)G(s-v),
\label{eq:sec:multifrac_0180}
\end{equation}
and
\begin{eqnarray}
  \sigma^2(t) & = \int_0^t\int_0^t G(u)C(u-v)G(v)dudv  \nonumber \\
              & = 2\int_0^tG(u)\int_0^u C(u-v)G(v)dudv  .
\label{eq:sec:multifrac_0190}
\end{eqnarray}
Assuming $t > s$, (\ref{eq:sec:multifrac_0180}) 
becomes
\begin{equation}
  K(s,t) = \int_0^sdv\Bigl[G(s-v)G_C(t-v) + G(t-v)G_C(s-v)\Bigr],
\label{eq:sec:multifrac_0200}
\end{equation}
with
\begin{equation}
  G_C(t) = (G*C)(t) = \int_0^t G(t-u)C(u).
\label{eq:sec:multifrac_0210}
\end{equation}

Here we would like to remark that throughout this paper we assume the stochastic processes under consideration have zero means, 
so the mean squared displacement is equal to variance. 
These two terms will be used interchangeably in our subsequently discussion.

Note that for the simplest case of 
(\ref{eq:sec:multifrac_0010}) 
with $m=1$, $n=1$ and $\alpha_1 = \alpha$, $\gamma_1 =1$, 
then for Riemann-Liouville (or Caputo) fractional derivative,
\begin{equation}
  D^\alpha x(t) = \eta(t),
\label{eq:sec:multifrac_0220}
\end{equation}
where $\xi_1(t) = \eta(t)$ is white noise.
For $D^{\alpha-1}x(t)\Bigr|_{t=0} = 0$
(or $x(0) = 0$),
(\ref{eq:sec:multifrac_0220}) 
defines Riemann-Liouville fractional Brownian motion or type II fractional Brownian motion with Hurst index $H$, $\alpha = H + 1/2$ \cite{Lim01}. 
The solution of (\ref{eq:sec:multifrac_0220}) 
with the above boundary condition is given by
\begin{equation}
  x(t) = I^\alpha\eta(t) = \frac{1}{\Gamma(\alpha)}\int_0^t (t - u)^{\alpha - 1} \eta(u) du ,
\label{eq:sec:multifrac_0230}
\end{equation}
with variance
\begin{equation}
  \sigma^2(t) = \frac{t^{2\alpha-1}}{(2\alpha -1)\left(\Gamma(\alpha)\right)^2}
              = \frac{t^{2H}}{2H\left(\Gamma(H+1/2)\right)^2} .
\label{eq:sec:multifrac_0240}
\end{equation}
Note that fractional Brownian motion of Riemann-Liouville type has a variance with the same time dependence as the standard fractional Brownian motion. In contrast to the later, 
though it is self-similar but its increment process is not stationary. 
However it has the advantage that the process begins at time $t = 0$,
and the Hurst index can take any value $H > 0$ 
\cite{Lim01}. 

Another simple case is when $m=1$, $n=1$ and $\gamma_1 = \gamma \neq 1$, 
which leads to a Gaussian non-stationary mono-fractal process with variance
\begin{equation}
  \sigma^2(t) = \frac{t^{2\alpha-\gamma}}{\sin(\gamma\pi/2)(2\alpha -\gamma)\Gamma(\alpha)\Gamma(\alpha -\gamma +1)} .
\label{eq:sec:multifrac_0250}
\end{equation}
This process can be subdiffusion or superdiffusion, depends on whether $2\alpha - \gamma < 1$ or $2\alpha - \gamma > 1$.

In the subsequent sections we shall consider various specific cases of (\ref{eq:sec:multifrac_0010}) 
for modeling accelerating and retarding anomalous diffusion.


\section{Accelerating Anomalous Diffusion}
\label{sec:accelerating}
One of the simplest model for accelerating diffusion can be obtained by using a special case of 
(\ref{eq:sec:multifrac_0010}):
\begin{equation}
  D^\alpha x(t) = \sum_{j=1}^n c_j \xi_{\gamma_j}.
\label{eq:accelerating_0010}
\end{equation}
The Green function is given by $G(t) = \frac{t^{\alpha-1}}{\Gamma(\alpha)}$ and
\begin{eqnarray}
  G_C(t) & = \int_0^t du \frac{(t-u)^{\alpha-1}}{\Gamma(\alpha)}\left[\sum_{j=1}^n c_j^2 \frac{u^{-\gamma_j}}{2\sin(\pi\gamma_j/2)\Gamma(1-\gamma_j)}\right] \nonumber \\
        & = \sum_{j=1}^n c_j^2 \frac{u^{\alpha-\gamma_j}}{2\sin(\pi\gamma_j/2)\Gamma(\alpha-\gamma_j+1)} .
\label{eq:accelerating_0020}
\end{eqnarray}
The covariance is given by
\begin{equation}
  K(s,t) = \sum_{j=1}^n K_j(s,t) ,
\label{eq:accelerating_0030}
\end{equation}
and from (\ref{eq:sec:multifrac_0200}) 
one gets
\begin{subequations}
\begin{eqnarray}\fl
 K_j(s,t)  = \int_0^s \frac{c_j^2}{2\sin(\pi\gamma_j/2)}
               \Biggr[ 
                  \frac{(s-u)^{\alpha-1}}{\Gamma(\alpha)}
                  \frac{(t-u)^{\alpha-\gamma_j}}{\Gamma(\alpha-\gamma_j+1)} \nonumber \\
           +
                  \frac{(t-u)^{\alpha-1}}{\Gamma(\alpha)}
                  \frac{(s-u)^{\alpha-\gamma_j}}{\Gamma(\alpha-\gamma_j+1)}
               \Biggl]
               du                
\label{eq:accelerating_0040a}
\\
\fl \qquad\quad
           = \frac{c_j^2}{2\sin(\pi\gamma_j/2)}
               \Biggr[ 
                  \frac{s^\alpha t^{\alpha-\gamma_j}}{\Gamma(\alpha)\Gamma(\alpha-\gamma_j+1)} 
                  \int_0^1 du(1 -u)^{\alpha-1}\biggl(1 - \frac{s}{t}u^{\alpha-\gamma_j}\biggr)^{\alpha-\gamma_j}
                  \nonumber \\
                 +
                  \frac{s^{\alpha-\gamma_j+1} t^{\alpha-1}}{\Gamma(\alpha)\Gamma(\alpha-\gamma_j+1)} 
                  \int_0^1 du (1 -u)^{\alpha-1}\biggl(1 - \frac{s}{t}u^{\alpha-\gamma_j}\biggr)^{\alpha-\gamma_j} 
               \Biggl]
                    \nonumber \\
\fl \qquad\quad
           = \frac{c_j^2}{2\sin(\pi\gamma_j/2)}
               \Biggr[ 
                  \frac{s^\alpha t^{\alpha-\gamma_j}}{\Gamma(\alpha+1)\Gamma(\alpha-\gamma_j+1)} 
                  F(\gamma_j-\alpha,1,1+\alpha,s/t)
                  \nonumber \\
                 +
                  \frac{s^{\alpha-\gamma_j+1} t^{\alpha-1}}{\Gamma(\alpha)\Gamma(\alpha-\gamma_j+2)} 
                  F(1-\alpha,1,2+\alpha-\gamma_j,s/t)
               \Biggl] ,
\label{eq:accelerating_0040b}
\end{eqnarray}
\end{subequations}
where we have used 
\cite{GradshteynRyzhik94}, 
\#9.111, page 1005. 
By using the following identity:
\begin{equation}
  F(\alpha,\beta,\gamma,1) = \frac{\Gamma(\gamma)\Gamma(\gamma-\alpha-\beta)}{\Gamma(\gamma-\alpha)\Gamma(\gamma-\beta)},
\label{eq:accelerating_0050}
\end{equation}
one gets the variance as
\begin{equation}\fl\qquad
  \sigma^2(t) = \sum_{j=1}^n\sigma_j^2(t) = \sum_{j=1}^n K_j(t,t)
              = \sum_{j=1}^n \frac{c_j^2t^{2\alpha-\gamma_j}}{(2\alpha-\gamma_j)\sin(\pi\gamma_j/2)\Gamma(\alpha)\Gamma(\alpha-\gamma_j+1)} .
\label{eq:accelerating_0060}
\end{equation}
Note that the variance can also be obtained directly from 
(\ref{eq:accelerating_0040a}). 

\begin{figure}[t]
  \centering
  \includegraphics[width=0.75\textwidth]{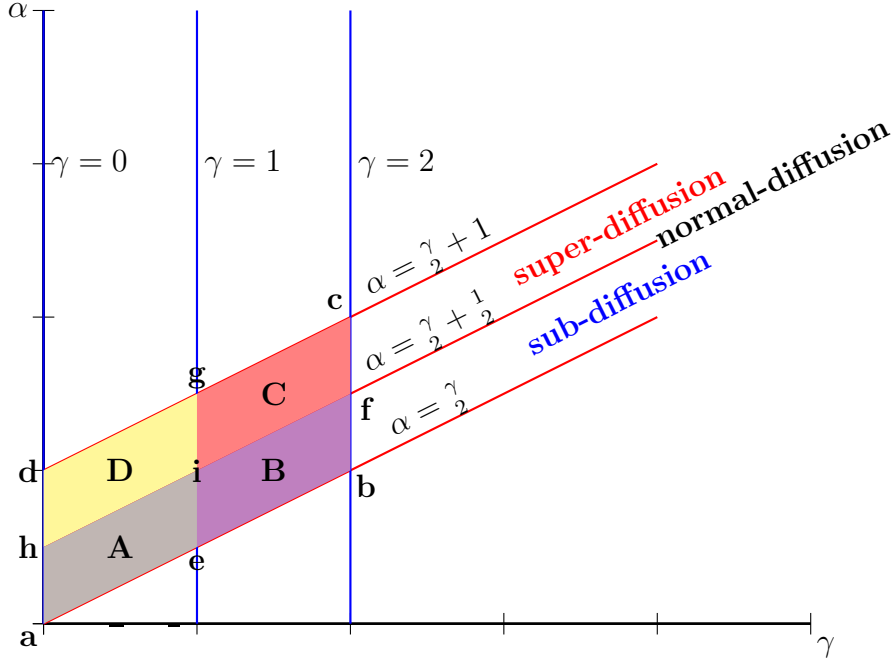}
  \caption{\label{fig:accelerating_0010}The domain of different anomalous diffusion types}
\end{figure}

For the case with $n = 2$, and $\gamma_1 > \gamma_2$, 
then the short-time and long-time limits of the mean-square displacement (or variance) are given by
\begin{equation}
  \sigma^2(t) \sim 
     \left\{
       \begin{array}{ll} \displaystyle
         \frac{c_1^2t^{2\alpha-\gamma_1}}{\sin(\pi\gamma_1/2)(2\alpha-\gamma_1)\Gamma(\alpha)\Gamma(\alpha-\gamma_1+1)} & \textrm{as} \ t \to 0   \\
         \displaystyle
         \frac{c_2^2t^{2\alpha-\gamma_2}}{\sin(\pi\gamma_2/2)(2\alpha-\gamma_2)\Gamma(\alpha)\Gamma(\alpha-\gamma_2+1)} & \textrm{as} \ t \to \infty
       \end{array}
     \right. .
\label{eq:accelerating_0080}
\end{equation}
Therefore the process is accelerating subdiffusion if for $i =1,2$,
$0 < (2\alpha -\gamma_i) < 1$, and it becomes accelerating superdiffusion when $1 < (2\alpha -\gamma_i) < 2$.
For $0 < (2\alpha -\gamma_1) < 1$ and $1 < (2\alpha -\gamma_2) < 3$,
the process begins as a subdiffusion and it accelerates to become a superdiffusion. 
Figure {\bf \ref{fig:accelerating_0010}}  shows the various possible type of anomalous diffusion for different values of $\alpha$ and $\gamma$.

In order to have some idea about the type of process $x(t)$ represents, 
let us consider some specific examples correspond to certain given values for $\alpha$ and $\gamma_i$.
For (\ref{eq:accelerating_0010}) 
with $\alpha=1$, the $i$th component of covariance of the process associated with the solution becomes
\begin{eqnarray}
  K_i(s,t) & = \frac{c_i^2}{
             2\sin(\pi\gamma_i/2)
               \Gamma(2-\gamma_i)}\int_0^s\Bigl[(t-u)^{1-\gamma_i} + (t-u)^{1-\gamma_i}\Bigr]du  \nonumber \\
           & = \frac{c_i^2}{
             2\sin(\pi\gamma_i/2)
             \Gamma(3-\gamma_i)}\Bigl[t^{2-\gamma_i} + s^{2-\gamma_i} - (t-s)^{2-\gamma_i}\Bigr],
\label{eq:accelerating_0090}
\end{eqnarray}
which is the covariance of the fractional Brownian motion (up to a multiplicative constant). 
Thus the covariance of the process $K(s,t) = \sum_{i=1}^n K_i(s,t)$ is the covariance of a mixed fractional Brownian motion (also called fractional mixed fractional Brownian motion by some authors) 
\cite{Cheridito01,ElNouty03,Thale09}
which is the sum of $n$ independent fractional Brownian motion
\begin{equation}
  x(t) = \sum_{i=1}^n c_iB_{H_i}(t),
\label{eq:accelerating_0100}
\end{equation}
where $H_i = (2-\gamma_i)/2$ is the Hurst index of the fractional Brownian motion $B_{H_i}(t)$.
Thus, in the mixed fractional Brownian motion model, anomalous diffusion begins with a lower diffusion rate can be represented by the fractional Brownian motion of lower Hurst index, 
and it is subsequently accelerated and is described by fractional Brownian motion of higher Hurst index.
It is interesting to note if all $H_i$ are not equal to $\frac{1}{2}$ (that is when they are all not Brownian motion), such a process is long-range dependent 
\cite{DoukhanOppenheimTaqqu03,LimMuniandy03}. 
The process satisfies a generalization of self-similar property called mixed self-similarity in the following sense:
\begin{equation}
  \sum_{i=1}^n B_{H_i}(rt) \eqind \sum_{i=1}^n r^{H_1}B_{H_i}(t).
\label{eq:accelerating_0110}
\end{equation}
The variance of the process is given by
\begin{equation}
  \sigma^2(t) = \sum_{j=1}^n \frac{c_j^2 t^{2H_j}}{
             \sin(\pi{H_j})
             \Gamma(2H_j +1)} .
\label{eq:accelerating_0120}
\end{equation}
Just like the previous case, suffice to consider $n=2$.
For $H_2 > H_1$,
the short and long time limits of the MSD are
\begin{equation}
  \sigma^2(t) \sim \left\{\!
                    \begin{array}{ll} \displaystyle
                      \frac{c_1^2t^{2H_1}}{
                          \sin(\pi{H_1})
                        \Gamma(2H_1+1)} & \textrm{as} \ t \to 0, \\[0.5cm]
                      \displaystyle
                      \frac{c_1^2t^{2H_1}}{
                          \sin(\pi{H_2})
                        \Gamma(2H_2+1)} & \textrm{as} \ t \to \infty.
                      \end{array}
                    \right.
\label{eq:accelerating_0130}
\end{equation}
Thus the process behaves as accelerating superdiffusion (or subdiffusion) for $1/2 < H_j < 1$ (or $0 < H_j < 1/2$) with $j =1,2$.

Here we would like to remark that there exists another process called step fractional Brownian motion 
\cite{BenassiBertrandCohenIstas00,AyacheBertrandLevyVehel07,LimTeo09} 
which can also be used to model both accelerating and retarding anomalous diffusion. Although the mixed fractional Brownian motion can only be used to describe accelerating anomalous diffusion, its mathematical structure is comparatively simpler than that of step fractional Brownian motion.

Next, we consider another special case with the exponents $\alpha$ and $\gamma_i$ satisfy the condition $\alpha-\gamma_i=0$ for a particular ith fractional Gaussian noise. 
(\ref{eq:accelerating_0040a}) 
now becomes
\begin{eqnarray}
  K_i(s,t) & = \frac{c_i^2}{
                 2\sin(\pi\alpha/2)
               \Gamma(\alpha)} 
             \int_0^s \Bigl[(s-u)^{\alpha-1} + (t-u)^{\alpha-1}\Bigr]du \nonumber \\
          & = \frac{c_i^2}{
                 2\sin(\pi\alpha/2)
               \Gamma(\alpha+1)} 
              \Bigl[s^\alpha + t^\alpha - (t-s)^\alpha\Bigr].
\label{eq:accelerating_0140}
\end{eqnarray}
Since $\gamma_i = 2 -2H_i=\alpha$, 
the process with the above covariance is fractional Brownian motion indexed by $2-2H_i$.
Note that the other components of covariance $K_j(s,t)$, $j \neq i$ are given by 
(\ref{eq:accelerating_0040b}) 
hence they are not fractional Brownian motion. The variance is given by
\begin{eqnarray} 
  \sigma^2(t) & = \sigma_i^2(t) + \sum_{j=1,j\neq i}^n \sigma_j^2(t) \nonumber \\
              & = \frac{c_i^2t^\alpha}{
                    2\sin(\pi\alpha/2)
                  \Gamma(\alpha+1)}   \nonumber \\
              &\quad  + \sum_{j=1,j\neq i}^n  \frac{c_j^2t^{2\alpha-\gamma_j}}{
                    2\sin(\pi\gamma_j/2)
                  (2\alpha-\gamma_j)
                  \Gamma(\alpha)\Gamma(\alpha-\gamma_j+1)}.
\label{eq:accelerating_0150}
\end{eqnarray}
In this case the type of anomalous diffusion depends on values of $\alpha$ and $2\alpha - \gamma_j$.
Assuming  $0 < \gamma_j < 1$, then for $0 < \alpha < 1$ and $0 < 2\alpha - \gamma_j < 1$ one gets accelerating subdiffusion;  
when $1 < \alpha < 2$ and $1 < 2\alpha - \gamma_j < 2$, 
the process is accelerating superdiffusion. 
The process accelerates (retards) from subdiffusion (superdiffusion) to become superdiffusion (subdiffusion) when $1 < \alpha < 2$ and $0 < 2\alpha - \gamma_j <1$ (or when $0 < \alpha < 1$ and $1 < 2\alpha - \gamma_j < 2$).

Finally, we note that if $\xi_i$ is white noise, that is when $\gamma_i = 1$, then $K_i(s,t)$ takes the following form:
\begin{equation}
  K_i(s,t) = \frac{c_j^2s^{\alpha} t^{\alpha-1}}{\Gamma(\alpha+1)\Gamma(\alpha)}
             F(1-\alpha,1,1+\alpha,s/t),
\label{eq:accelerating_0160}
\end{equation}
which is just the covariance of ``type II'' or Riemann-Liouville fractional Brownian motion 
\cite{Lim01}. 
The variance for the process is given by
\begin{equation}
  \sigma_i^2(t) = \frac{2c_i^2t^{2\alpha-1}}{(2\alpha-1)(\Gamma(\alpha))^2} .
\label{eq:accelerating_0170}
\end{equation}
For $n=2$, $\gamma_1 =1$, and $\gamma_2 < 1$,
one gets a process which is an accelerating subdiffusion if $1/2 < \alpha < 1$;
an accelerating superdiffusion if $1 < \alpha < 3/2$,
and finally it represents a process which accelerates from a subdiffusion to a superdiffusion if $1/2 \alpha < 1$ and $\gamma_2 < 2\alpha -1$.

From the above discussion it is noted that a simple fractional Langevin equation driven by a single fractional Gaussian noise results in an anomalous diffusion. 
However, when the process is driven by more than single fractional Gaussian noise term, 
the resulting process can be accelerating subdiffusion or superdiffusion, 
depending on the fractional order of the noise terms and the derivative term.  
Thus, the interplay between multiple driving fractional Gaussian noise terms in the fractional Langevin-like equation leads to a simple Gaussian model of accelerating anomalous diffusion.


\section{Retarding Anomalous Diffusion}
\label{sec:retarding}
There exists another type of anomalous diffusion that slows down with time. 
In other words, for short times the mean square displacement of such a diffusion process varies as $t^{\alpha_1}$,
and it varies as $t^{\alpha_2}$ with $\alpha_2 < \alpha_1$
for long times.
Retarding anomalous diffusion occurs in various physical and biological systems. 
For examples, in the single-file diffusion that occurs in cell membranes and narrow channels 
\cite{VestergaardBogind85,WeiBechingerLeiderer00}, 
and the retarding anomalous diffusion in semicondutors
\cite{TanGosele05,Servidori87}. 

A simple Gaussian model for retarding anomalous diffusion can be obtained by considering a special case of
(\ref{eq:sec:multifrac_0010}) 
which takes the form of following fractional stochastic differential equation:
\begin{equation}
  a_1D^{\alpha_1}x(t) + a_2D^{\alpha_2}x(t) = \sum_{h=1}^n c_j \xi_{\gamma_j}(t), \qquad 0 < \alpha_2 < \alpha_1 < 2 .
\label{eq:retarding_0010}
\end{equation}
For both Riemann-Liouville and Caputo case, the Laplace transform of 
(\ref{eq:retarding_0010}) 
gives
\begin{equation}
  \Bigl[a_1s^{\alpha_1} + a_2s^{\alpha_2}\Bigr]\tilde{x}(s) - B(s) = \sum_{j=1}^n c_j \xi_{\gamma_j}(t) ,
\label{eq:retarding_0020}
\end{equation}
where $B(s)$ is given by (\ref{eq:sec:multifrac_0150}). 
For simplicity, we again choose the initial conditions such that $B(s) = 0$. 
The Green function is then given by the inverse Laplace transform of
\begin{equation}
  \widetilde{G}(s) = \frac{1}{a_1s^{\alpha_1} + a_2s^{\alpha_2}} = \frac{1}{a_1} \frac{s^{-\alpha_2}}{s^{\alpha_1-\alpha_2} + (a_2/a_1)}
\label{eq:retarding_0030}
\end{equation}
such that its inverse Laplace transform gives
\begin{equation}
  G(t) = \frac{1}{a_1} t^{\alpha_1-1} E_{\alpha_1-\alpha_2,\alpha_1}\left(-\frac{a_2}{a_1}t^{\alpha_1-\alpha_2}\right) ,
\label{eq:retarding_0040}
\end{equation}
where
\begin{equation}
  E_{\mu,\nu}(z) = \sum_{j=0}^\infty \frac{z^j}{\Gamma({\mu}j + \nu)}, \qquad \mu > 0, \nu > 0,
\label{eq:retarding_0050}
\end{equation}
is the Mittag-Leffler function \cite{ErdelyiIII55}.
Using (\ref{eq:sec:multifrac_0200}) 
and 
(\ref{eq:sec:multifrac_0210}), 
and substituting $C(t) = \sum_{j=1}^n c_j^2 \frac{t^{-\gamma_j}}{\Gamma(1-\gamma_j)}$ one gets
\begin{eqnarray}
  G_C(t) & = \sum_{j=1}^n c_j^2 \int_0^t du \frac{1}{a_1}(t-u)^{\alpha_1-1} 
             E_{\alpha_1-\alpha_2,\alpha_1}\left(-\frac{a_2}{a_1}(t-u)^{\alpha_1-\alpha_2}\right)
             \frac{u^{-\gamma_j}}{\Gamma(1-\gamma_j)} \nonumber \\
         & = \sum_{j=1}^n \frac{c_j^2}{a_1}t^{\alpha_1-\gamma_j} 
             E_{\alpha_1-\alpha_2,\alpha_1-\gamma_j +1}\left(-\frac{a_2}{a_1}t^{\alpha_1-\alpha_2}\right).
\label{eq:retarding_0060}
\end{eqnarray}
The covariance $K(s,t) = \sum_{j=1}^n c_j^2 K_j(s,t)$ is given by 
(\ref{eq:sec:multifrac_0200}) 
with
\begin{eqnarray}\fl
  K_j(s,t) = \int_0^s du
               \Biggl[
                 \frac{(s-u)^{\alpha_1-1}}{a_1}
                 E_{\alpha_1-\alpha_2,\alpha_1}\left(-\frac{a_2}{a_1}(s-u)^{\alpha_1-\alpha_2}\right)
               \Biggr]  \nonumber \\
               \qquad
               \Biggl[
                 \frac{(t-u)^{\alpha_1-\gamma_j}}{a_1}
                 E_{\alpha_1-\alpha_2,\alpha_1-\gamma_j+1}\left(-\frac{a_2}{a_1}(t-u)^{\alpha_1-\alpha_2}\right)
               \Biggr]    \nonumber \\
           + \int_0^s du
               \Biggl[
                 \frac{(t-u)^{\alpha_1-1}}{a_1}
                 E_{\alpha_1-\alpha_2,\alpha_1}\left(-\frac{a_2}{a_1}(t-u)^{\alpha_1-\alpha_2}\right)
               \Biggr]  \nonumber \\
               \qquad\qquad
               \Biggl[
                 \frac{(s-u)^{\alpha_1-\gamma_j}}{a_1}
                 E_{\alpha_1-\alpha_2,\alpha_1-\gamma_j+1}\left(-\frac{a_2}{a_1}(s-u)^{\alpha_1-\alpha_2}\right)
               \Biggr] .
\label{eq:retarding_0070}
\end{eqnarray}
The variance is given by $\sigma^2(t) = \sum_{j=1}^n c_j^2 K_j(t,t)$
\begin{eqnarray}
  K_j(t,t) =  2\int_0^s du
               \Biggl[
                 \frac{(t-u)^{\alpha_1-\gamma_j-1}}{a_1^2}
                 E_{\alpha_1-\alpha_2,\alpha_1}\left(-\frac{a_2}{a_1}(t-u)^{\alpha_1-\alpha_2}\right)
                 \nonumber \\
                 \qquad\qquad\qquad\qquad\qquad\qquad
                 E_{\alpha_1-\alpha_2,\alpha_1-\gamma_j+1}\left(-\frac{a_2}{a_1}(t-u)^{\alpha_1-\alpha_2}\right)
               \Biggr] .
\label{eq:retarding_0080}
\end{eqnarray}

Note that the covariance and variance given above can not be evaluated. However, by using the following asymptotic properties of Mittag-Leffler function 
\cite{ErdelyiIII55}, 
\begin{equation}\fl
  E_{\mu,\nu}(-z) \sim - \sum_{n=1}^N \frac{(-1)^{n-1}z^{-n}}{\Gamma(\nu - n\mu)} + \Or\Bigl(|z|^{-1-N}\Bigr), \quad |\arg(z)| < \left(1-\frac{\mu}{2}\right)\pi, \ z \to \infty,
\label{eq:retarding_0090}
\end{equation}
and
\begin{equation}
  E_{\mu,\nu}(-z) \sim \frac{1}{\Gamma(\nu)} + \Or(z), \quad z \to 0 ,
\label{eq:retarding_0100}
\end{equation}
it is possible to obtain the short and long time behavior of the variance. 
In the case of single fractional Gaussian noise, one has for $\gamma < 2\alpha_2$,
\begin{equation}
  \sigma^2(t) \sim \left\{\!
                     \begin{array}{ll}
                       \displaystyle
                       t^{2\alpha_1-\gamma}, & t \to 0  \\
                       \displaystyle
                       t^{2\alpha_2-\gamma}, & t \to \infty
                     \end{array}
                   \right. .
\label{eq:retarding_0110}
\end{equation}
Since $\alpha_1 > \alpha_2$, the process is a retarding subdiffusion (or superdiffusion) if for $i=1,2$, $0 < \alpha_i -\gamma/2 < 1/2$ (or $1/2 < \alpha_i -\gamma/2 < 1$).
In the case when there are more than one noise, 
the dominant terms for the short time limit and long time limit for the variance are $\sim t^{\min(2\alpha_1-\gamma_j)}$ and  $\sim t^{\max(2\alpha_2-\gamma_j)}$ respectively. 
Thus we see that the double-order fractional stochastic equation driven by single fractional Gaussian noise can be used to model retarding anomalous diffusion. 
The lower order fractional derivative term in (\ref{eq:retarding_0010}) 
plays the role of a damping term which slows down the diffusion.

The main disadvantage of using (\ref{eq:retarding_0010}) 
for modeling retarding anomalous diffusion is that in general the covariance and variance of the underlying process can not be calculated explicitly. 
We would like to find a particular case of 
(\ref{eq:retarding_0010}) 
such that its solution is a process with covariance and variance that can be completely determined. 
For this purpose we consider the following double-order fractional Langevin-like equation:
\begin{equation}
  a_1D^{\alpha_1}x(t) + a_2D^{\alpha_2}x(t) = c_1\xi_1(t) + c_2\xi_2(t), \qquad 0 < \alpha_2 < \alpha_1 < 2.
\label{eq:retarding_0120}
\end{equation}
The two independent fractional Gaussian noises $\xi_1(t)$ and $\xi_2(t)$ are chosen such that for $t > s$ they have zero mean and the following covariance
\begin{equation}
  \Bigl<\xi_i(t)\xi_j(s)\Bigr> = \frac{(t-s)^{\nu-\alpha_i-1}}{\Gamma(\nu-\alpha_i)} \delta_{ij} 
                               \equiv C_i(t-s)\delta_{ij}, \quad i,j = 1,2 .
\label{eq:retarding_0130}
\end{equation}
We remark that the fractional Gaussian noise with covariance given by
{(\ref{eq:retarding_0130})} 
is selected based on practical purposes as it gives the required results as well as provides a more manageable solution to 
(\ref{eq:retarding_0120}). 
As a result 
(\ref{eq:retarding_0130}), 
one can verify that
the Laplace transforms of the covariance of the fractional Gaussian noise $C(t) = c_1^2C_1(t) + c_2^2C_2(t)$ and the Green function  $G(t)$ satisfy the following relation:
\begin{equation}
  \widetilde{G}(s)\widetilde{C}(s) = s^{-\nu}.
\label{eq:retarding_0140}
\end{equation}
The inverse Laplace transform of (\ref{eq:retarding_0140}) 
is
\begin{equation}
  G(t)*C(t) = \frac{t^{\nu-1}}{\Gamma(\nu)} .
\label{eq:retarding_0150}
\end{equation}
The covariance of the process associated with the solution of (\ref{eq:retarding_0120}) 
can be calculated using (\ref{eq:sec:multifrac_0200}) 
and  (\ref{eq:retarding_0150}): 
\begin{eqnarray} \fl
  K(s,t)  = \int_0^s du\Biggl[G(t-u)\frac{(s-u)^{\nu-1}}{\Gamma(\nu)} + G(s-u)\frac{(t-u)^{\nu-1}}{\Gamma(\nu)}\Biggr]  \nonumber \\
\fl \qquad\quad
          = \int_0^s du\Biggl[
             \frac{(t-u)^{\alpha_1-1}}{a_1}E_{\alpha_1-\alpha_2,\alpha_1}\left(-\frac{a_2}{a_1}(t-u)^{\alpha_1-\alpha_2}\right)
             \frac{(s-u)^{\nu-1}}{\Gamma(\nu)} \nonumber \\         
          \qquad\qquad   + \frac{(s-u)^{\alpha_1-1}}{a_1}E_{\alpha_1-\alpha_2,\alpha_1}\left(-\frac{a_2}{a_1}(s-u)^{\alpha_1-\alpha_2}\right)\frac{(t-u)^{\nu-1}}{\Gamma(\nu)}
             \Biggr]  .
\label{eq:retarding_0160}
\end{eqnarray}
Its variance is given by the following series expansion:
\begin{eqnarray}
  \sigma^2(t) & = 2 \int_0^t du \Biggr[
             \frac{1}{a_1\Gamma(\nu)}(t-u)^{\alpha_1+\nu-2}E_{\alpha_1-\alpha_2,\alpha_1}\left(-\frac{a_2}{a_1}(t-u)^{\alpha_1-\alpha_2}\right)
             \Biggl]   \nonumber \\
             & = \frac{2}{a_1\Gamma(\nu)}
                 \sum_{n=0}^\infty \frac{\left(\frac{-a_2}{a_1}\right)^n t^{n(\alpha_1-\alpha_2)+\alpha_1+\nu-1}}
                                     {(n(\alpha_1-\alpha_2)+\alpha_1+\nu-1)\Gamma(n(\alpha_1-\alpha_2)+\alpha_1)} .
\label{eq:retarding_0170}
\end{eqnarray}
From (\ref{eq:retarding_0170}) 
one gets the short and long time limits of the variance as
\begin{equation}
  \sigma^2(t) \sim \left\{
                  \begin{array}{ll}
                    \displaystyle
                    \frac{2}{a_1\Gamma(\nu)\Gamma(\alpha_1)\Gamma(\alpha_1+\nu-1)}t^{\alpha_1+\nu-1}, & t \to 0 \\
                    \displaystyle
                    \frac{2}{a_1\Gamma(\nu)\Gamma(\alpha_2)\Gamma(\alpha_2+\nu-1)}t^{\alpha_2+\nu-1}, & t \to \infty
                  \end{array}
                \right. .
\label{eq:retarding_0180}
\end{equation}
The process with covariance (\ref{eq:retarding_0170}) 
and variance (\ref{eq:retarding_0180}) 
is a retarding subdiffusion or superdiffusion depending on the on the values of $\alpha_i + \nu -1$, $i = 1,2$.
Thus, one can regard the term $a_2D^{\alpha_2}x(t)$ in (\ref{eq:retarding_0120}) 
as a damping term which slows down the anomalous diffusion.

A special case for which the covariance has a closed form is when $\nu=1$.
For $t > s$, the covariance  (\ref{eq:retarding_0160}) 
becomes
\begin{eqnarray} \fl
  K(t,s) =  \int_0^s du 
            \Bigg[
            \frac{1}{a_1}(t-u)^{\alpha_1-1} E_{\alpha_1-\alpha_2,\alpha_1}\left(-\frac{a_2}{a_1}(t-u)^{\alpha_1-\alpha_2}\right)
            \nonumber \\
            +
            \frac{1}{a_1}(s-u)^{\alpha_1-1} E_{\alpha_1-\alpha_2,\alpha_1}\left(-\frac{a_2}{a_1}(s-u)^{\alpha_1-\alpha_2}\right)
            \Bigg]  \nonumber \\
\fl \qquad\quad
         = \frac{t^{\alpha_1}}{a_1} E_{\alpha_1-\alpha_2,\alpha_1+1}\left(-\frac{a_2}{a_1}t^{\alpha_1-\alpha_2}\right)
         + \frac{s^{\alpha_1}}{a_1} E_{\alpha_1-\alpha_2,\alpha_1+1}\left(-\frac{a_2}{a_1}s^{\alpha_1-\alpha_2}\right) \nonumber \\
         - \frac{(t-s)^{\alpha_1}}{a_1} E_{\alpha_1-\alpha_2,\alpha_1+1}\left(-\frac{a_2}{a_1}(t-s)^{\alpha_1-\alpha_2}\right) .
\label{eq:retarding_0190}
\end{eqnarray}
It is interesting to note that the covariance consists of three terms of Mittag-Leffler functions of same order, 
and the time variables $t$ and $s$ enter 
the covariance expression {(\ref{eq:retarding_0190})} 
in a form similar to that of fractional Brownian motion.
Thus it is not a coincidence that the asymptotic short and long time limits of the covariance are given by
\begin{equation}
  K(t,s) = \left\{
  \begin{array}{ll}
    \displaystyle
    \frac{1}{a_1\Gamma(\alpha_1+1)} \Bigl[t^{\alpha_1} + s^{\alpha_1} - |t-s|^{\alpha_1}\Bigr], & t,s \to 0 \\
        \displaystyle
    \frac{1}{a_2\Gamma(\alpha_2+1)} \Bigl[t^{\alpha_2} + s^{\alpha_2} - |t-s|^{\alpha_2}\Bigr], & t,s, |t-s| \to \infty
  \end{array}\right. .
\label{eq:retarding_0200}
\end{equation}
Note that these are just the covariance of fractional Brownian motion indexed respectively by $\alpha_1/2 = 1- H_1$ and $\alpha_2/2 = 1- H_2$, with $0 < H_i < 1$, $i = 1,2$.
Since both the short and long time limits are fractional Brownian motion, 
one can conclude that the stochastic process in this case is long-range dependent except when $\alpha_1 = \alpha_2 =1/2$ or when both the limiting processes are Brownian motion 
\cite{DoukhanOppenheimTaqqu03,LimMuniandy03}.

The variance is given by
\begin{eqnarray}
  \sigma^2(t) & = \frac{2}{a_1}
                t^{\alpha_1} 
                E_{\alpha_1-\alpha_2,\alpha_1+1}\left(-\frac{a_2}{a_1}t^{\alpha_1-\alpha_2}\right)  .
\label{eq:retarding_0210}
\end{eqnarray}
The long and short time limits are given by
\begin{equation}
  \sigma^2(t) \sim \left\{\!\!
                   \begin{array}{ll}
                     \displaystyle
                     \frac{2}{a_1\Gamma(\alpha_1+1)}t^{\alpha_1} , & t \to 0 \\
                     \displaystyle
                     \frac{2}{a_2\Gamma(\alpha_2+1)}t^{\alpha_2} , & t \to \infty
                   \end{array}
               \right. .
\label{eq:retarding_0220}
\end{equation}
From the above results one has for $0<\alpha_2 < \alpha_1 <1$ a subdiffusion process which slows down with time, 
or a retarding subdiffusion. 
On the other hand, if $1<\alpha_2 < \alpha_1 <2$,
the process is a retarding superdiffusion.

It would be interesting to see whether (\ref{eq:retarding_0120}) 
can be used to describe accelerating anomalous diffusion as well. 
Suppose we replace the condition (\ref{eq:retarding_0140}) 
by the following:
\begin{equation}
  \widetilde{G}(s)\widetilde{C}(s) = s^{-\nu} + s^{-\kappa} ,
\label{eq:retarding_0230}
\end{equation}
such that
\begin{equation}
  \widetilde{C}(s) = a_1s^{\alpha_1-\nu} + a_1s^{\alpha_1-\kappa}
                     a_2s^{\alpha_2-\nu} + a_2s^{\alpha_2-\kappa} .
\label{eq:retarding_0240}
\end{equation}
Inverse Laplace transform of  (\ref{eq:retarding_0240}) 
gives
\begin{equation}
  C(t) = a_1\Biggl[
                  \frac{t^{\nu-\alpha_1-1}}{\Gamma(\nu-\alpha_1)}
                  + \frac{t^{\kappa-\alpha_1-1}}{\Gamma(\kappa-\alpha_1)}
            \Biggr]
            +
            a_2\Biggl[
                  \frac{t^{\nu-\alpha_2-1}}{\Gamma(\nu-\alpha_2)}
                  + \frac{t^{\kappa-\alpha_2-1}}{\Gamma(\kappa-\alpha_2)}
            \Biggr].
\label{eq:retarding_0250}
\end{equation}
The variance of the resulting process is
\begin{equation}
  \sigma^(t) = \sigma_\nu^2(t) + \sigma_\kappa^2(t) ,
\label{eq:retarding_0260}
\end{equation}
with $\sigma_\nu^2(t)$ given by (\ref{eq:retarding_0170}) 
and similarly for $\sigma_\kappa^2(t)$ with $\nu$ replaced by $\kappa$.
Thus for $\nu < \kappa$, one gets
\begin{equation}
  \sigma^2(t) \sim \left\{\!
                   \begin{array}{ll}
                     \displaystyle
                     \frac{2}{a_1\Gamma(\nu)\Gamma(\alpha_1)\Gamma(\alpha_1+\nu-1)}t^{\alpha_1+\nu-1}, & t \to 0 \\[0.5cm]
                     \displaystyle
                     \frac{2}{a_2\Gamma(\kappa)\Gamma(\alpha_2)\Gamma(\alpha_2+\kappa-1)}t^{\alpha_2+\kappa-1}, & t \to \infty
                   \end{array}
                   \right. .  
\label{eq:retarding_0270}
\end{equation}
It is interesting to note that the process with the above variance represents accelerating subdiffusion if $\alpha_1+\nu < \alpha_2 + \kappa$,
or $\kappa > \alpha_1 -\alpha_2 +\nu > 2\alpha_1-\alpha_2$.
However, to achieve such an accelerating subdiffusion it is necessary to consider four fractional 
Gaussian noise terms in (\ref{eq:retarding_0120}), 
a situation that may be difficult to realize in practice.

On the other hand, we recall that for (\ref{eq:retarding_0010}) 
with more than one fractional Gaussian noise, 
the dominant term of the variance for the associated process in the short and long time limit is respectively varies as $t^{min(2\alpha_1-\gamma_j)}$ and $t^{max(2\alpha_2-\gamma_j)}$.
If we consider the case with $\alpha_1 > \alpha_2$ and $\gamma_1 > \gamma_2$,
one then has $2\alpha_1-\gamma_2 > 2\alpha_1 -\gamma_1$ and $2\alpha_2-\gamma_2 > 2\alpha_2 -\gamma_1$. 
As a result,
\begin{equation}
  \sigma^2(t) \sim \left\{
                  \begin{array}{ll}
                    \displaystyle
                    t^{2\alpha_1-\gamma_1}, & t \to 0  \\
                    \displaystyle
                    t^{2\alpha_2-\gamma_2}, & t \to \infty
                  \end{array}
                \right. .    
\label{eq:retarding_0280}
\end{equation}
Thus, it is possible to use (\ref{eq:retarding_0120}) 
to model accelerating anomalous diffusion provided $2\alpha_2 -\gamma_2 > 2\alpha_1 -\gamma_1$ or $\alpha_1-\alpha_2 < (\gamma_1 -\gamma_2)/2$.
In other words, the fractional stochastic equation (\ref{eq:retarding_0120}) 
can be used to model retarding and accelerating anomalous diffusion by appropriate choice of the order of the fractional derivatives and fractional Gaussian noise terms.
Note that retarding diffusion such as single-file diffusion can also be modeled by fractional generalized langevin equation
\cite{EabLim10}. 


\section{Concluding remarks}
\label{sec:conclude}
We have shown that it is possible to model both accelerating and retarding anomalous diffusion by using fractional Langevin-like stochastic differential equations driven by one or more terms of fractional Gaussian noise. 
The solutions associated with some specific cases of these equations turn out to be some interesting processes in the short and long time limits. 
For example, two types of fractional Brownian motion, namely the usual standard fractional Brownian motion and the Riemann-Liouville fractional Brownian motion are the asymptotic processes of special cases of the model. 
This model also includes another interesting process,
namely the mixed fractional Brownian motion, 
which is a simple process for describing accelerating sub- and super-diffusion.

We note that the stochastic differential equations in our model can be regarded as fractional Langevin-like equation of distributed order 
(\ref{eq:sec:multifrac_0100}) 
with weight function consists of delta functions.
One may want to consider cases with different type of weight functions in 
(\ref{eq:sec:multifrac_0100}), 
such as uniform or power-law weight functions. 
However, from the results of our previous study on fractional Langevin equations of distributed order with uniform and power-law type of weight functions indicates that such equations in general do not have closed solutions even for the case of simple fractional Langevin of distributed order driven by white noise
\cite{EabLim11}. 
One thus expects the situation to be even more complex when weight functions other than the delta functions are used for the multi-term fractional Langevin equation with more than one fractional noise terms.

One question of interest is that whether it is possible to model accelerating and retarding anomalous diffusion based on 
(\ref{eq:accelerating_0010}) 
and (\ref{eq:retarding_0010}), 
using different Gaussian noise. 
If one is only interested in the asymptotic limits of the mean squared displacement of the stochastic process, 
then instead of using fractional Gaussian noise in the stochastic differential equations 
(\ref{eq:accelerating_0010}) 
and 
(\ref{eq:retarding_0010}), 
Gaussian noise with covariance which has the ``correct'' asymptotic limits of power-law type can be used. For example, Gaussian noise with covariance which varies as $At^{\mu-2}/\left(1+At^{\mu-\nu}/B\right)$,
$A$ and $B$ are positive constants, $0 < \mu,\nu < 2$.
Such a covariance has respectively short and long time limit $At^{\mu-2}$ and $Bt^{\nu-2}$ respectively. 
Another example is the noise of Mittag-Leffler type with covariance that of the form ${At^{\mu-2}} E_{\mu-\nu,\mu+1}\left(-At^{\mu-\nu}/B\right)$, $\mu > \nu$, 
which has short and long time asymptotic limit $At^{\mu-2}/\Gamma(\mu+1)$ and $Bt^{\nu-2}/\Gamma(\nu+1)$ respectively.
These two examples give the possible alternatives to the fractional Gaussian noise $\xi_{\gamma_j}$ 
as the driving noise.
More discussion related to these cases will be given in a forthcoming paper 
\cite{EabLimxx}. 

The usual characterization of anomalous diffusion uses its mean squared displacement (or variance in the context of this paper).
However, it is well-known that even for a Gaussian model mean squared displacement does not determine completely the underlying stochastic process and hence the mechanism of the anomalous diffusion. 
Recent advances in particle tracking devices allow experiments to track the trajectories of single molecule or nanoparticle in complex systems such as cells in a biological system.
Mean squared displacement obtained from the time series data gives the scaling exponent of the anomalous diffusion undertaken by such particles. 
Comparison of the experimental data so obtained with various models of anomalous diffusion allows one to distinguish the different possible subdiffusion mechanisms. 
In particular, information of single-particle trajectories allows one to test the validity of ergodic property of the associated diffusion process. 
A stochastic process is said to be ergodic if the ensemble average of certain physical quantity such as mean squared displacement measured in bulk coincides with the time average of the same quantity over sufficiently long time from the single-molecule time series. 
Examples of ergodic process are Brownian motion and the standard fractional Brownian motion.  
Another process of interest which is ergodic is mixed fractional Brownian motion which is the sum of two independent fractional Brownian motion. 
On the other hand, Riemann-Liouville fractional Brownian motion 
and heavy tailed continuous-time random walk are non-ergodic
\cite{BelBarkai05,DengBarkai09,Fulinski11,BurovJeonMetzlerBarkai11}.

By using the single-molecule data the comparisons of experimental data based on various models of anomalous diffusion such as the continuous-time random walk, fractional Brownian motion, fractional Levy stable motion, etc. have been carried out by various authors recently 
\cite{Wong04,SzymanskiWeiss09,Tejedor10,Jeonetal11,GoldingCox06,%
MagdziarzWeronBurneckiKlafter09,BurneckiWeron10,KeptenBronshteinGarini11}.
For examples, diffusion of beads in entangled F-actin networks, diffusion of at shorter times exhibits continuous-time random walk behaviour 
\cite{Wong04}.
However, the analysis of the data of the anomalous diffusion in crowded intracellular fluid such as cytoplasm of living cells rules out continuous-time random walk and favours fractional Brownian motion 
\cite{SzymanskiWeiss09}.
Analysis of single particle tracking data of lipid granules in yeast cells by Tejedor et al 
\cite{Tejedor10}
seems to rule out continuous-time random walk and shows agreement with fractional Brownian motion; 
but a subsequent study 
\cite{Jeonetal11}
shows that at short times the granules perform continuous-time random walk subdiffussion while at longer times the motion is consistent with fractional Brownian motion. 
Various analyses of the biological data describing the motion of individual fluorescently labeled mRNA molecules inside live E. coli cells, a well-known experiment first conducted by Golding and Cox 
\cite{GoldingCox06}
do not lead to consensus result. 
Magdziarz et al 
\cite{MagdziarzWeronBurneckiKlafter09}
show that fractional Brownian motion as the underlying stochastic process; 
but subsequently Burnecki and Weron claim that the data follows fractional Levy stable motion 
\cite{BurneckiWeron10}.
According to Kepten et al the experiments on telomeres
in the nucleus of the mammalian cell exhibit fractional Brownian motion 
\cite{KeptenBronshteinGarini11}.
Recent study by A.V.Weigel et al 
\cite{Weigel11}
on the physical mechanism underlying Kv2.1 voltage gated potassium channel anomalous dynamics using single-molecule tracking  shows that both ergodic (diffusion on a fractal) and nonergodic (continuous-time random walk) processes coexist in the plasma membrane.   
Though it is widely recognized that the diffusion pattern of membrane protein displays anomalous subdiffusion, 
however, there is still no agreement on the mechanisms responsible for this transport behaviour. 
Currently there is still no consensus on whether heavy tailed continuous time random walk, fractional Brownian motion, fractional Levy stable motion or some other stochastic processes can provide the correct description to anomalous diffusion in some biological systems.
Thus, it is important to make use of data and information other than the mean squared displacement or the anomalous diffusion exponent to determine the type of mechanism and the stochastic process describing the anomalous diffusion. 
We hope that some of the processes considered in this paper may be of relevance in describing the anomalous diffusion in biological systems.

Finally we remark that it would be interesting to investigate the Fokker-Planck equations associated with the processes considered in this paper. 
The mean first passage time for some of the simpler cases can also be studied
\cite{EabLimxx}. 


\ack{%
S. C. Lim would like to thank the Malaysian Ministry of Science, Technology and Innovation and Malaysian Academy of Sciences for the support under the Brain Gain Malaysia (Back to Lab) Program.
He would also like to thank Eli Barkai and Ralf Metzler for their useful comments on stochastic processes and mechanisms describing anomalous diffusion in biological systems.
}

\section*{References}
\bibliographystyle{iopart-num}
\bibliography{biblioAD}


\end{document}